\documentclass[smallcondensed]{svjour3}     % onecolumn (ditto)
\usepackage{graphicx}
\usepackage{lineno,hyperref}
\usepackage{graphicx}  
\usepackage{subfigure}
\usepackage{amsfonts}
\usepackage{color}
\usepackage{float}
\usepackage{booktabs}
\usepackage{multirow}
\usepackage[numbers]{natbib}
\usepackage{paralist}
\usepackage{algorithm}
\usepackage{algorithmic}

\begin{document}

\title{Heterogeneous Information Network-based Interest Composition with Graph Neural Network for Recommendation}

\titlerunning{HicRec}        % if too long for running head

\author{Dengcheng Yan  \and
        Wenxin Xie     \and
		Yiwen Zhang
}


\institute{Dengcheng Yan \at
              School of Computer Science and Technology, Anhui University, Hefei, China \\
              \email{yanzhou@ahu.edu.cn}           
           \and
           Wenxin Xie \at
              School of Computer Science and Technology, Anhui University, Hefei, China\\
				\email{xiewxahu@foxmail.com}
		\and
		Yiwen Zhang \at
			School of Computer Science and Technology, Anhui University, Hefei, China\\
				\email{zhangyiwen@ahu.edu.cn}
}


\maketitle

\begin{abstract}

Heterogeneous information networks (HINs) are widely applied to recommendation systems due to their capability of modeling various auxiliary information with meta-paths. However, existing HIN-based recommendation models usually fuse the information from various meta-paths by simple weighted sum or concatenation, which limits performance improvement because it lacks the capability of interest compositions among meta-paths. In this article, we propose an \textbf{H}IN-based \textbf{I}nterest \textbf{C}omposition model for \textbf{Rec}ommendation (HicRec). Specifically, user and item representations are learned with a graph neural network on both the graph structure and features in each meta-path, and a parameter sharing mechanism is utilized here to ensure that the user and item representations are in the same latent space. Then, users' interests in each item from each pair of related meta-paths are calculated by a combination of the user and item representations. The composed user interests are obtained by their single interest from both intra- and inter-meta-paths for recommendation. Extensive experiments are conducted on three real-world datasets and the results demonstrate that our proposed HicRec model outperforms the baselines.

.
\end{abstract}

\section{Introduction} \label{intro}

Recommendation systems (RS) address the information overload problem, which overwhelms users in the era of information explosion and is widely applied in various fields such as production recommendations in e-commerce \cite{e_commerce_rec, rs_app1} and service recommendations \cite{service_rec, rs_app3}. Generally, RS models users' historical interactions with items as a bipartite network and learns users' preferences by similarity measures \cite{CF} or latent models \cite{MF}. However, these models usually {face} the challenge of data sparsity. To alleviate this challenge, various auxiliary information such as social relations \cite{social} and contents \cite{rs_app2} have been used to enhance RS performance. Recently, heterogeneous information network (HIN) \cite{hin} was proposed to model this auxiliary information, including diversified entities and multiple relations in a unified framework. Moreover, meta-path, a powerful analysis tool for HIN, is designed to capture rich semantics between nodes to ensure the resilience of HIN-based RS against data sparsity.
 
\begin{figure}
\subfigure[Intra-meta-path]{
\includegraphics[scale=0.25]{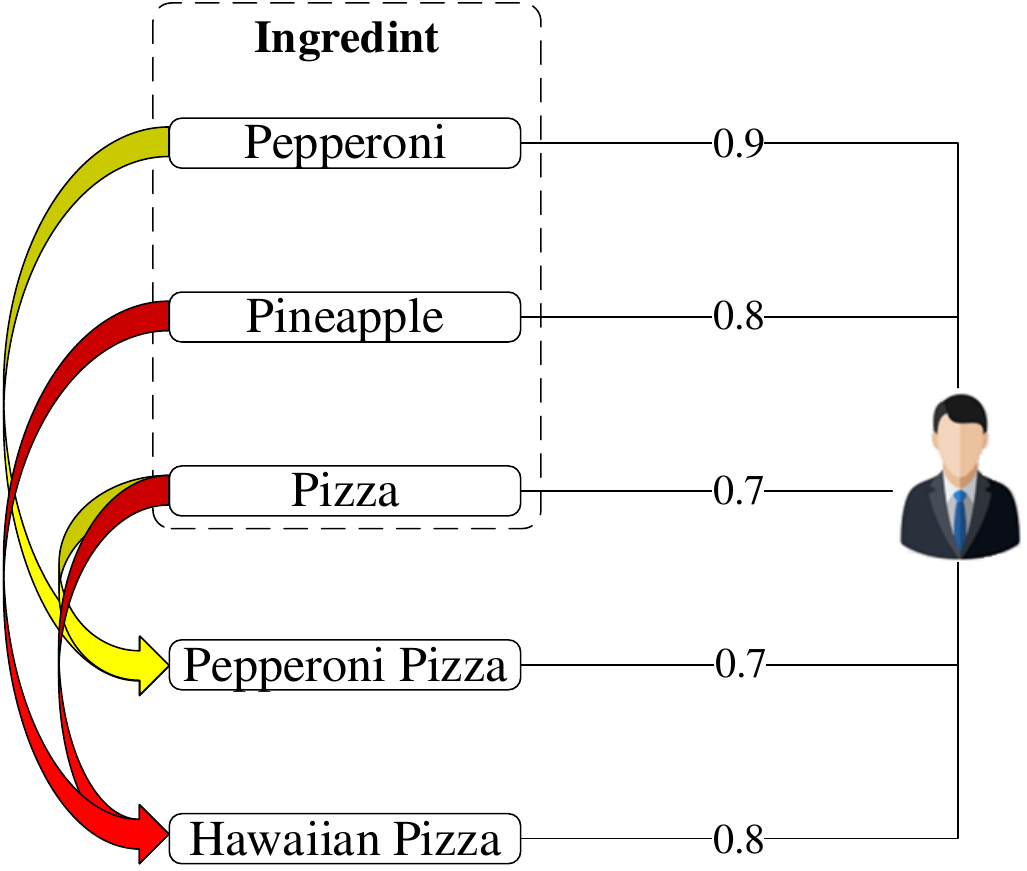}
\label{intra}
}
\subfigure[Inter-meta-paths]{
\includegraphics[scale=0.25]{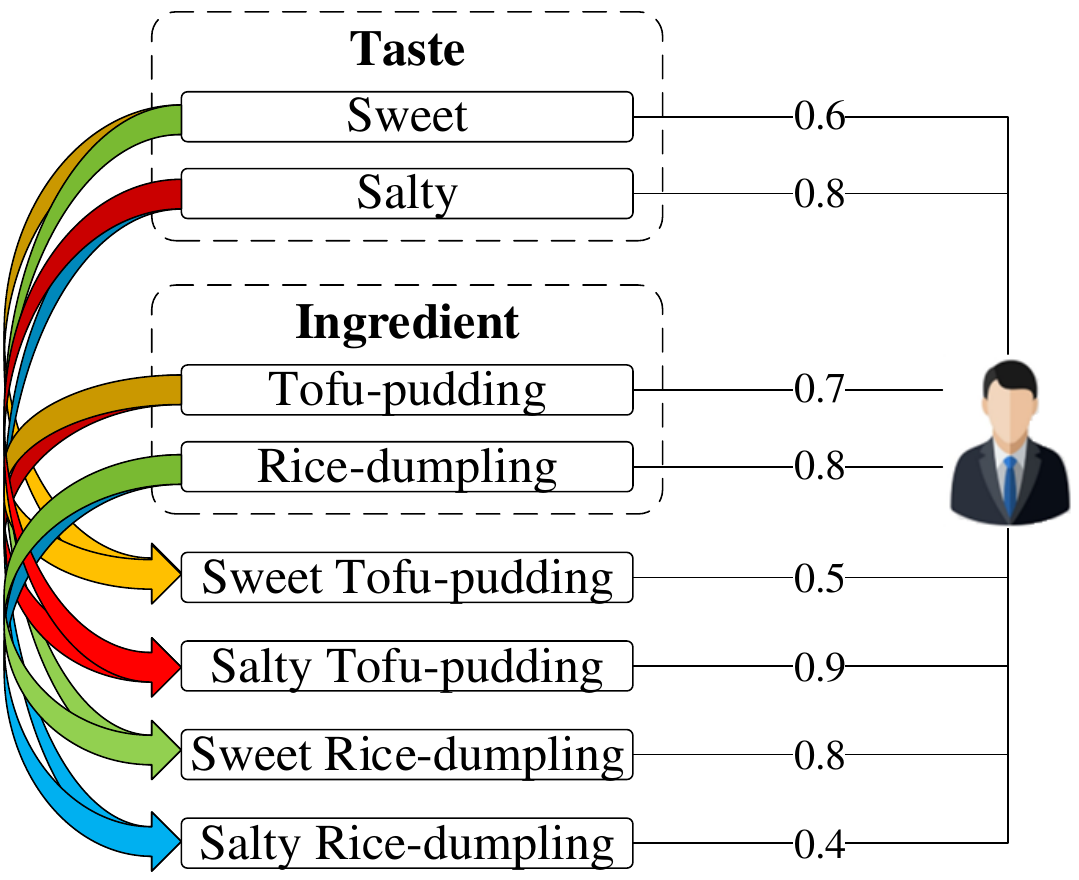}
\label{inter}
}
\caption{Examples of interest composition}
\label{ic_exmaple}
\end{figure}

In the HIN-based RS framework \cite{heterec, fmg, neuacf}, there generally exist two essential stages, i.e., information extraction from each meta-path and information fusion from multiple meta-paths for recommendation. In the information extraction stage, multiple HIN subgraphs are extracted using predefined meta-paths and the latent user and item representations are then obtained from these subgraphs by methods such as matrix factorization and multilayer perceptrons. Then in the information fusion stage, the extracted information from each meta-path is combined by methods such as pooling and attention mechanism \cite{attention}. This process merges information from multiple aspects for recommendation. However, there still exist some drawbacks to this framework. First, the subgraph structure and user/item features are tightly related while the information extraction stage seldom considers them jointly \cite{hnafm, neuacf, herec, hoprec}. Second, the information fusion stage usually ignores the higher-order feature composition of users' interests from either intra- or inter-meta-paths. Figure \ref{ic_exmaple} gives an example of interest composition from both intra- and inter-meta-path. As Figure \ref{intra} shows, within the user-ingredient meta-path the user's interest in different ingredients (i.e., pizza, pepperoni and pineapple) can be obtained from his/her historical behaviors, while his/her composited interest in Hawaiian pizza or pepperoni pizza is not a simple combination of interests for each ingredient. Similarly, as Figure \ref{inter} shows, interest can be composited from inter-meta-paths (i.e., user-taste and user-ingredient meta-paths) and the user's preference for the composited interests needs to be learned.

In this paper, we propose the \textbf{H}eterogeneous information network-based \textbf{i}nterest \textbf{c}omposition for \textbf{Rec}ommendation (HicRec) model. To improve the representation ability of user and item embedding in the information extraction stage, we introduce a graph neural network (GNN) to consider both the graph structure and user/item features from each meta-path. Moreover, the extracted information about users and items is further composited in the information fusion stage via interest composition to capture higher-order interests.

Our main contributions can be summarized as follows:
\begin{itemize}
	\item 
	We propose a novel HIN-based recommendation model, HicRec, which combines the representation learning of users' interest from each meta-path and interest composition from intra- and inter-meta-paths into a unified framework. A graph neural network is used in the information extraction stage for each meta-path in order to jointly consider both the graph structure and user/item features.

	\item
	We design two interest composition mechanisms to capture users’ high-order composited interests. Intra- and inter-meta-path interest compositions generate new composited interests by combining features within a single meta-path interest representation and across different meta-path interest representations.

	\item
	We conduct extensive experiments on three real-world datasets. The results demonstrate that our proposed HicRec outperforms the baselines.
\end{itemize}

The rest of the paper is organized as follows. Section \ref{related-work} briefly reviews the related works, and Section \ref{preliminaries} gives some preliminaries. In Section \ref{model}, our proposed HicRec model is described in detail. The experiments and analysis are presented in Section \ref{experiments}. Finally{, we give a conclusion}  in Section \ref{conclusions}.

\section{Related Work} \label{related-work}

\subsection{Recommendation Systems}
The majority of conventional {RS} are based on collaborative filtering (CF) \cite{CF}, which assumes that users with similar historical interactions should have similar preferences in the future. As one of the most common CF-based RS, matrix factorization (MF) \cite{MF} learns the low-rank embeddings of users and items by factorizing the user-item interaction matrix. Based on MF, various variants, such as {B}ayesian {P}ersonalized {R}anking (BPR) \cite{bpr} and {S}ingular {V}alue {D}ecomposition (SVD) \cite{svd} have been proposed. However, these simple factorization models are usually powerless in terms of representation ability and weak in capturing the complex interactions between users and items. Fortunately, with the prevalence of deep learning, many {RS} have attempted to mine deeper relations by replacing the matrix factorization model with some deep learning technologies, such as multilayer perceptron (MLP) \cite{dmf}. Furthermore, He et al. \cite{ncf} argued that the computation of predicted scores by inner product has some limitations and proposed the NeuMF to solve this problem by computing predicted scores with MLP instead of the inner product. These {RS} mentioned above have less consideration of the feature crosses, while DeepFM \cite{deepfm} has been proposed to enhance the capability of feature interactions with the combination of a neural network and the FM \cite{fm}. Recently, research on graph structure data has achieved great progress, and its powerful inference capability is applied to various tasks, including recommendation. For example, He et al. modeled the interactions between users and items as bipartite a graph and propose the NGCF \cite{ngcf} to combine the idea of CF and graph structure data.

\subsection{HIN-based Recommendation Systems}
Conventional {RS} based on CF usually suffer from the problem of data sparsity, and various auxiliary information has been utilized to alleviate this problem. Since HIN models several types of information, many {RS} based on HIN have been proposed. The major differences among the current HIN-based {RS} are mainly in information extraction and fusion stage. In the information extraction stage, diverse information has been mined with various techniques. For example, matrix factorization is utilized in HeteRec \cite{heterec}, FMG \cite{fmg} and HueRec \cite{huerec}, while HNAFM \cite{hnafm} and NeuACF \cite{neuacf} adopt MLP to extract deeper user and item information according to their initialized features in each meta-path. In addition, HERec \cite{herec} and HopRec \cite{hoprec} capture the graph structure in the corresponding meta-path with the help of DeepWalk \cite{deepwalk}. Generally, these methods do not consider both graph structure and features jointly. In the information fusion stage, HIN-based {RS} usually merge information extracted from each meta-path and make recommendations according to the representations after fusion. Several operations have been used in this stage, such as concatenation and MLP. However, these methods treat all meta-paths equally and ignore that the preferences on different meta-paths are significantly different. Therefore, the attention mechanism \cite{attention} is adopted in some models, such as HNAFM and NeuACF, to fuse information in a more personalized way.

\subsection{Graph Neural Network}
Through applying the neural network to graph data, the GNN is a powerful method for mapping nodes in a graph into low-rank representations and is utilized for various downstream tasks. As the most classic GNN, the graph convolutional network (GCN) \cite{gcn} conducts information aggregation according to the adjacent matrix of graphs. However, the neighbors of a node may contribute differently in the information aggregation process. Thus, the graph attention network (GAT) \cite{gat} adopts the attention mechanism to aggregate information adaptively for nodes. Moreover, since the GCN is {transductive}, it needs to be retrained once the graph structure changes. Therefore, GraphSAGE \cite{graphsage} was proposed to learn node embeddings and inference in an inductive form. Different from the above methods based on homogeneous graphs, there are also some methods extending the GNN into HIN. For example, Wang et al. \cite{han} proposed the HAN model to apply the GAT to the corresponding graph in each meta-path and designed semantic-level attention to merge node embeddings in different meta-paths. Nevertheless, the meta-paths should be predefined in the HAN, and the selection of meta-paths determines the performance of the HAN. To settle this problem, the graph transformer network (GTN) \cite{gtn} is proposed to identify useful meta-paths in an HIN.

\section{Preliminaries} \label{preliminaries}

In this section, we give some definitions and examples of some basic HIN concepts.

\begin{figure}
	\subfigure[{Example of HIN and meta-path (highlighted with red)}]{
		\includegraphics[scale=0.40]{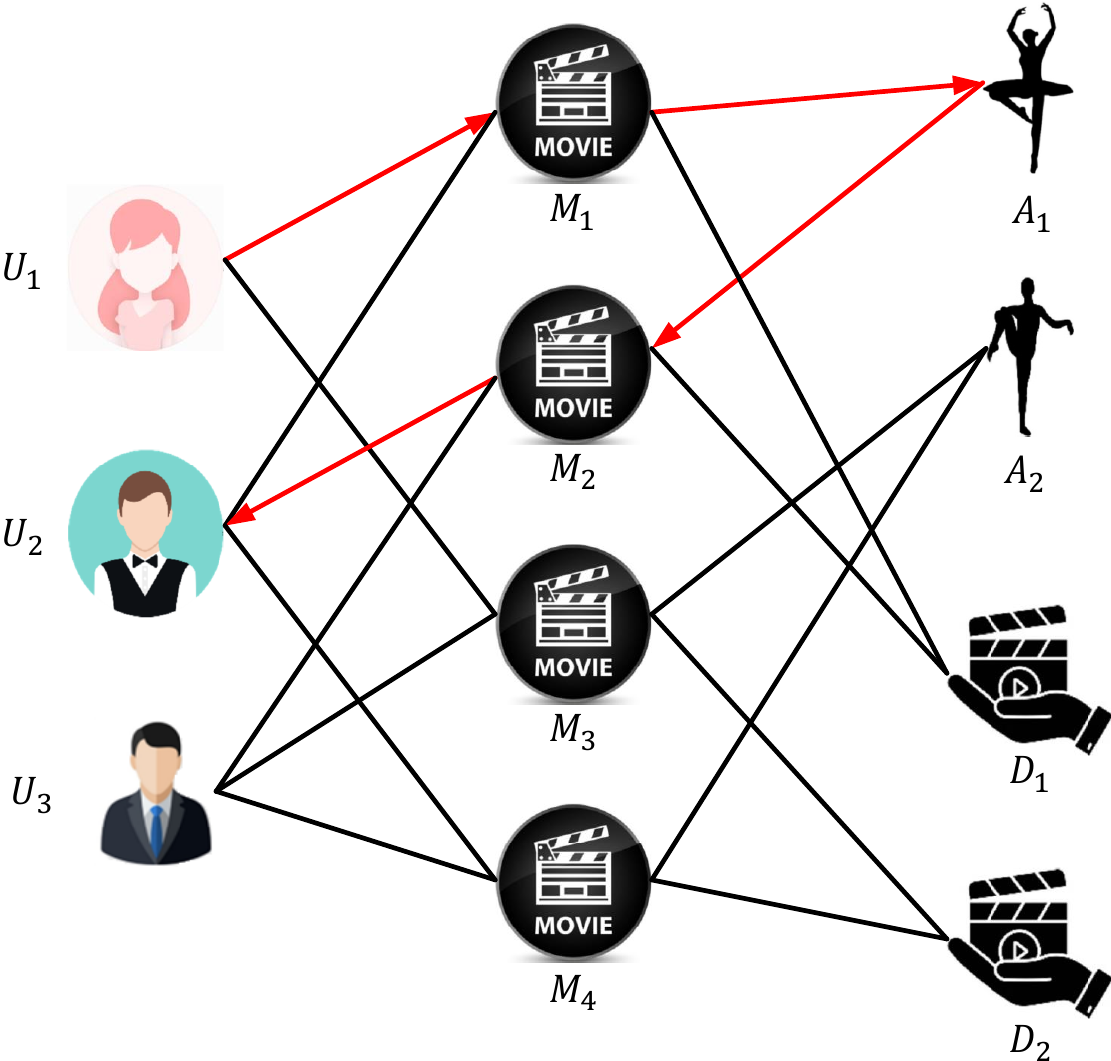}
		\label{hin}
	}
	\subfigure[Commuting matrix]{
		\includegraphics[scale=0.40]{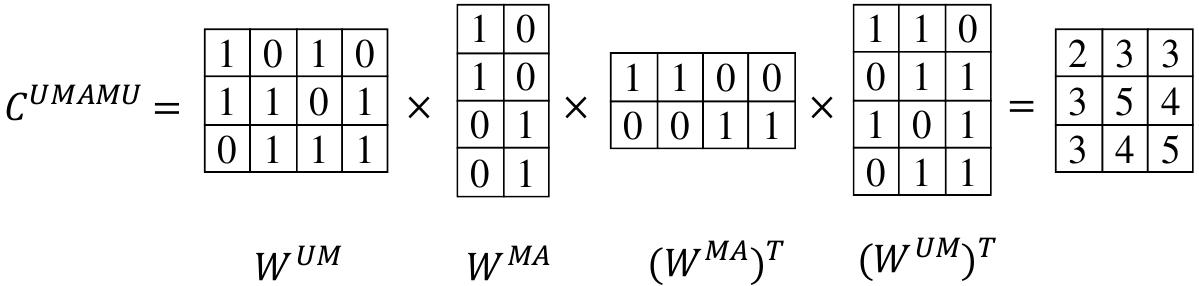}
		\label{cm}
	}
	\caption{{HIN, meta-path and commuting matrix}}
	\label{example}
\end{figure}

\noindent{\textbf{Definition 1 (HIN \cite{hin})}}: A heterogeneous information network is defined as a graph $G = (V; E)$, with a node type mapping function $\Phi :V \to A$ and a link type mapping function $\Psi :E \to R${, where} $A$ and $R$ are sets of node and link types, respectively. The restriction in $A$ and $R$ is: $\mid A \mid + \mid R \mid > 2$.

Figure \ref{hin} is an example of HIN on a movie viewing platform. As shown, there are multiple node types, such as actors and directors, as well as various link types, such as user-movie and movie-actor relations. Obviously, multitype information is quite useful for boosting recommendation performance. Thus, compared with conventional {RS}, which only model the relation between users and items, HIN-based {RS} can utilize more useful information for {recommendations}, especially for users with fewer history records.

\noindent{\textbf{Definition 2 (Meta-Path \cite{hin})}}: A meta-path $P$ is a path defined on a schema $S = (A,R)$, and is denoted in the form of $A_1\stackrel{R_1}{\longrightarrow}A_2\stackrel{R_2}{\longrightarrow}...\stackrel{R_l}{\longrightarrow}A_{l+1}$, which defines a composite relation $R=R_1\circ R_2\circ ...\circ R_l$ between objects $A_1, A_2,..., A_{l+1}$, where $\circ$ denotes the composition operator on relations.

Meta-path can express semantics between source nodes and target nodes. For example, in Figure \ref{hin}, $U_1-M_1-A_1-M_2-U_2$ is an instance of meta-path $UMAMU$, which means that both user $U_1$ and $U_2$ have seen the movie with actor $A_1$. In HIN-based {RS}, various meta-paths have been used to merge semantics into recommendations for more accurate results or explanations of RS \cite{explanation}.

\noindent{\textbf{Definition 3 (Commuting Matrix \cite{cm})}}: For a meta-path $P=(A_1 A_2…A_{l+1})$, the corresponding commuting matrix is defined as $\textbf{C}^P=\textbf{W}_{A_1 A_2}\textbf{W}_{A_2 A_3 }...\textbf{W}_{A_l A_{l+1}}$, where $\textbf{W}_{A_i A_{i+1}}$ is adjacent matrix between type $A_i$ and $A_{i+1}$.

As shown in Figure \ref{cm}, the commuting matrix for $UMAMU$ can reflect the strength of the relation between two users in terms of their preferences for common actors. For example, there are 3 paths from $U_1$ to $U_2$ with the meta-path schema of $UMAMU$ (i.e., $U_1 \to M_1 \to A_1 \to M_1 \to U_2$, $U_1 \to M_1 \to A_1 \to M_2 \to U_2$ and $U_1 \to M_3 \to A_2 \to M_4 \to U_2$). Thus, the link strength between $U_1$ and $U_2$ on the actor aspect is 3, which is represented by the corresponding value in commuting matrix $C^{UMAMU}$. Intuitively, a high value in the commuting matrix denotes the strong relation between two nodes.

\section{Proposed Model: HicRec} \label{model}

\subsection{Overview}
\begin{figure}
\includegraphics[scale=0.3]{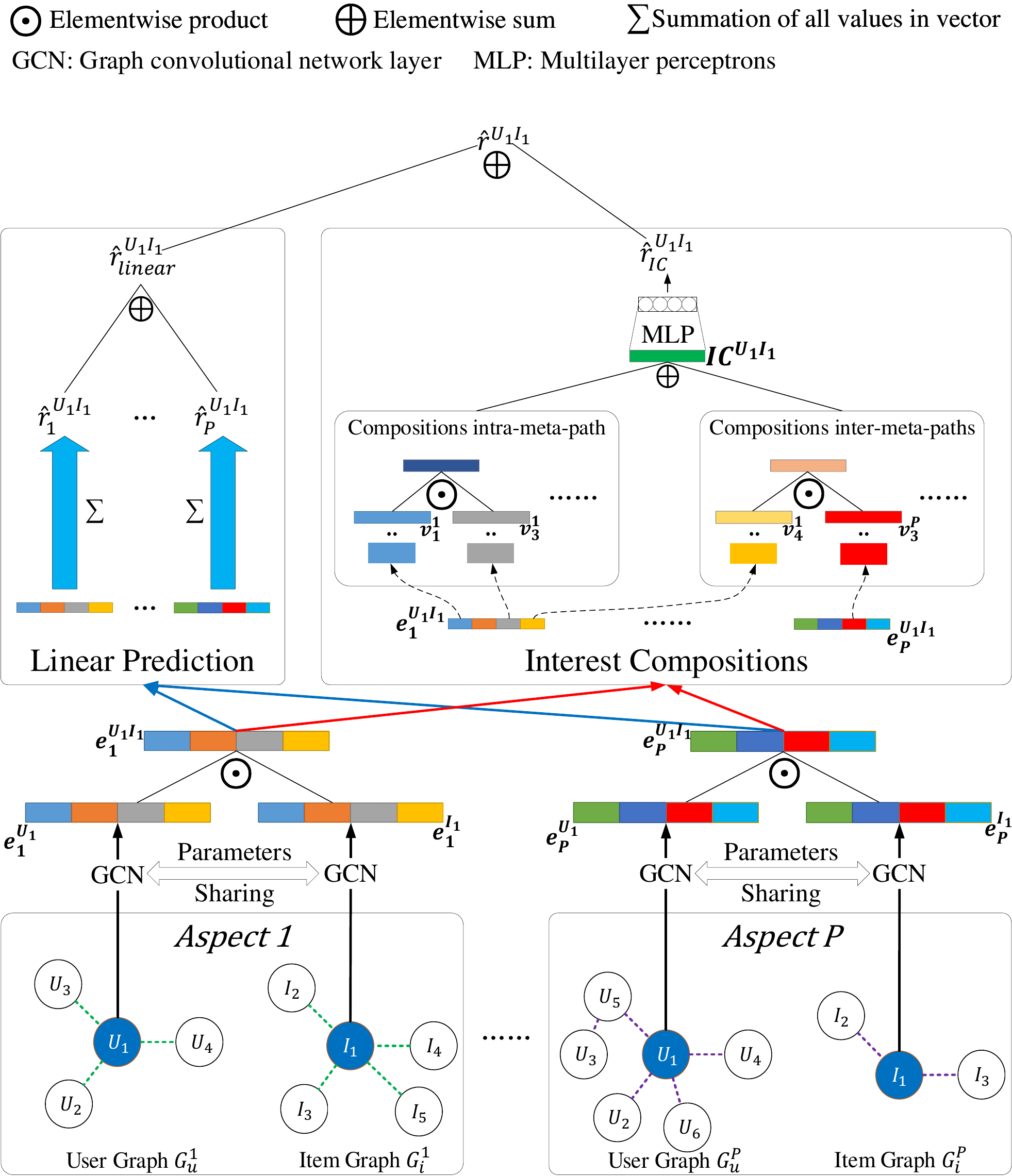}
\caption{{The overall framework of HicRec}}
\label{framework}
\end{figure}

In this section, we present the details of the proposed HicRec model, as shown in Figure \ref{framework}. Generally, HicRec consists of two steps, i.e., interest extraction and interest composition. In the graph corresponding to each meta-path, the interest extraction step {utilizes} the GCN \cite{gcn} to map users and items into low-rank embeddings by combining both the graph structure information and user/item features and the elementwise product is applied to transform the embeddings of each user-item pair into the user's interest on the item (Section \ref{interest-extraction}). Then we conduct both intra- and inter-meta-path interest composition with the help of neural factorization machine (NFM) \cite{nfm} (Section \ref{interest-composition}). Finally, the parameters in the HicRec model are updated according to the BPR loss for recommendation (Section \ref{parameter-learning}).

\subsection{Interest Extraction} \label{interest-extraction}

In this step, we adopt the GCN to extract information from meta-paths with the combination of graph structure and user/item features. First, we classify a pair of user- and item-related meta-paths about the same topic into one aspect \cite{neuacf}, and users' interests are extracted within each aspect. Taking Figure \ref{hin} as an example, user-related meta-path $UMAMU$ and item-related meta-path $MAM$ are both related to actors, representing users’ preferences for actors and the relations between movies and actors.

For each aspect $p$, we construct a user graph $G_u^p$ and an item graph $G_i^p$ by computing the commuting matrices of the pair of meta-paths as corresponding adjacent matrices. Then, the GCN is applied to the two graphs to obtain the embeddings of users and items as follows:

\begin{equation}
\textbf{U}_{p}^{l + 1} = \sigma(\textbf{A}_{G_{u}^{p}}\textbf{U}_{p}^{l}\textbf{W}^{l} + \textbf{b}^{l})
\label{user_gcn}
\end{equation}
\begin{equation}
\textbf{I}_{p}^{l + 1} = \sigma(\textbf{A}_{G_{i}^{p}}\textbf{I}_{p}^{l}\textbf{W}^{l} + \textbf{b}^{l})
\label{item_gcn}
\end{equation}
where $\textbf{U}_{p}^{l} \in \mathbb{R}^{\mid G_{u}^{p} \mid \times d}$ and $\textbf{I}_{p}^{l} \in \mathbb{R}^{\mid G_{i}^{p} \mid \times d}$ ($d$ is the embedding dimension, $\mid G_{u}^{p} \mid$ denotes the number of nodes in graph $G_{u}^{p}$) denote the $l$-th layer embeddings of users and items in aspect $p$, respectively. In addition, we utilize the similarity matrix calculated by PathSim \cite{cm} as the initialized features of users ($\textbf{U}_{p}^{0}$) and items ($\textbf{I}_{p}^{0}$) in aspect $p$, and {a nonlinear layer before the GCN is used to transform the similarity vectors of users and items into a uniform latent space}. $\textbf{A}_{G_u^p}$ and $\textbf{A}_{G_i^p}$ are the adjacent matrices of user graph $G_{u}^{p}$ and item graph $G_{i}^{p}$, respectively. $\textbf{W}^{l}\in \mathbb{R}^{d \times d}$ and $\textbf{b}^{l} \in \mathbb{R}^{d}$ are the parameters of $l$-th layer. It is worth noting that we adopt the mechanism of parameter sharing to ensure the embeddings of users and items in the same latent space. Moreover, we choose ReLU as the nonlinear activation function $\sigma$. Multiple layers of GCN can be stacked to capture the high-order graph structure information.

Due to the mechanism of parameter sharing, the user and item representations from the same aspect lie in the same latent space. Thus, the simple elementwise product is adopted to merge the user and item embeddings into the users' interests,
\begin{equation}
\textbf{e}^{ui}_{p} = \textbf{e}^{u}_{p} \odot \textbf{e}^{i}_{p}
\label{ewp}
\end{equation}
where $\textbf{e}_{p}^{u}$ is the $u$-th row of ${\textbf{U}_{p}^{L}}$ and $\textbf{e}_{p}^{i}$ is the $i$-th row of ${\textbf{I}_{p}^{L}}$. $\odot$ is the elementwise product operation.

\subsection{Interest Composition} \label{interest-composition}

Users' single interests from each aspect obtained from the previous step are still not sufficient for recommendation. Therefore, interest composition is needed to capture higher-order features. We design two interest composition mechanisms, intra- and inter-meta-path interest compositions, to generate new composited interests by combining features within a single meta-path interest representation and across different meta-path interest representations.

\paragraph{Intra-meta-path interest composition} is conducted by composing each pair of elements from a single interest representation $\textbf{e}_p^{ui}$ of a user in each aspect,
\begin{equation}
\textbf{intra}^{ui}_{p} = \sum_{m=1}^{d}\sum_{n=m+1}^{d}\textbf{e}^{ui}_{p, m}\textbf{v}^{p}_{m} \odot \textbf{e}^{ui}_{p, n}\textbf{v}^{p}_{n}
\label{equa_intra}
\end{equation}
where $\textbf{e}_{p, m}^{ui}$ denotes the $m$-th value in $\textbf{e}_p^{ui}$ and $\textbf{v}_{m}^p{}\in \mathbb{R}^K$ is the factor vector in aspect $p$ ($K$ is the dimension of {the} factor).

The intra-meta-path composition may not be enough to capture the complex patterns of user-item interactions. Actually, inter-meta-path interest composition is also important.

\paragraph{Inter-meta-path interest composition} is conducted by composing pairs of elements from different single interest representations,
\begin{equation}
\textbf{inter}^{ui} = \sum_{p=1}^{P}\sum_{q=p+1}^{P}\sum_{m=1}^{d}\sum_{n=1}^{d}\textbf{e}^{ui}_{p, m}\textbf{v}^{p}_{m} \odot \textbf{e}^{ui}_{q, n}\textbf{v}^{q}_{n}
\label{equa_inter}
\end{equation}
where $P$ is the total number of aspects.

After obtaining the composited interests $\textbf{intra}_p^{ui} \in \mathbb{R}^K$ and $\textbf{inter}^{ui} \in \mathbb{R}^K$ from both intra- and inter-meta-path separately, the overall interest can be calculated by their simple sum,
\begin{equation}
\textbf{IC}^{ui} = \textbf{inter}^{ui} + \sum^{P}_{p=1}\textbf{intra}_{p}^{ui}
\label{equation_ic}
\end{equation}

Finally, the predicted score can be obtained by a combination of linear prediction and interest composition,
\begin{equation}
\hat{r}_{u, i} = \sum_{p=1}^{P}\textbf{e}_{p}^{u}{\textbf{e}_{p}^{i}}^{T} + \textbf{W}_{Rec}^{T}\textbf{IC}^{ui}
\label{rate}
\end{equation}
where the $\textbf{W}_{Rec} \in \mathbb{R}^K$ is the parameter. Generally, the former part in this equation denotes linear prediction, while the latter can be regarded as high-order interaction beyond linear prediction.

\begin{figure}
\includegraphics[scale=0.3]{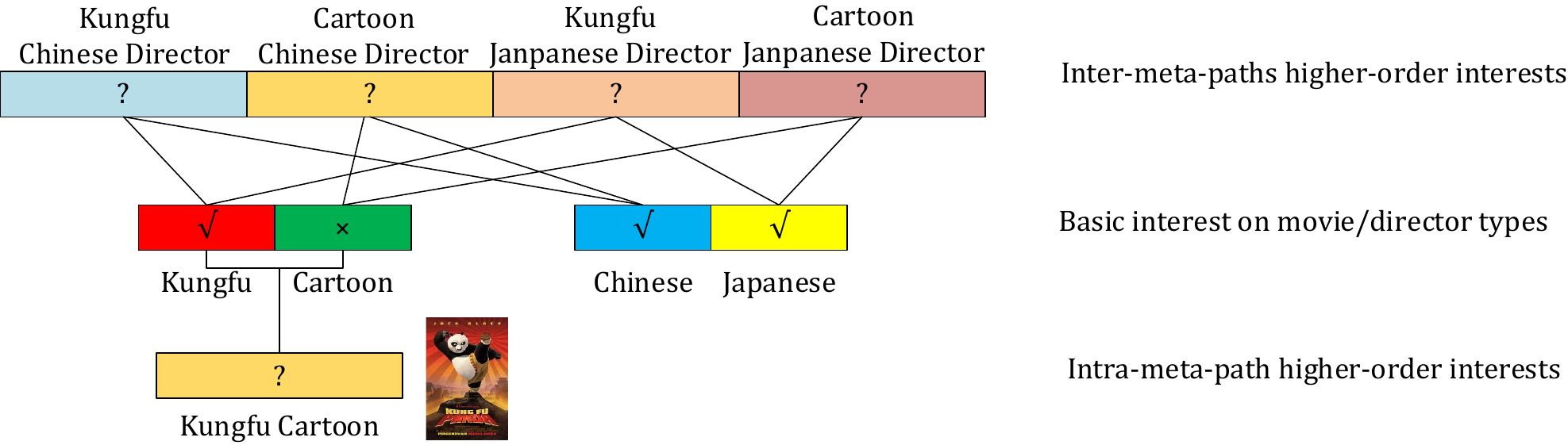}
\caption{{Explanation of intra- and inter-meta-paths compositions}}
\label{composition}
\end{figure}

{In the process of interest composition, users' basic interests extracted by GCN (i.e., $e_{p}^{ui}$) are explicitly composed by utilizing both intra- and inter- meta-paths interest composition for sufficient recommendation. Specially, intra-meta-path interest composition is conducted by composing each pair of the feature dimensions of a same basic interest, e.g., in Figure \ref{composition}, a new feature \textit{Kungfu Cartoon} is derived from the same basic interest on movie types. Inter-meta-path interest composition is conducted by composing each pair of the feature dimensions from two different basic interests, e.g., in Figure \ref{composition}, new features like  \textit{Kungfu Chinese Director}, \textit{Cartoon Japanese Director} and so on are derived from the two different basic interests on movie and director types. With the help of interest composition, more features are derived for sufficient recommendation.}

\begin{algorithm}
\caption{{Training of HicRec}}
\label{algorithm_hicrec}
\begin{algorithmic}[1]
\REQUIRE Set of aspects $\{1, 2, \cdots, P\}$, set of commuting matrices for each aspect $\{(\textbf{A}_{G_u^1}, \textbf{A}_{G_i^1}), (\textbf{A}_{G_u^2}, \textbf{A}_{G_i^2}), \cdots ,(\textbf{A}_{G_u^P}, \textbf{A}_{G_i^P})\}$, set of initialized user/item features for each aspect $\{\textbf{(U}_1^0, \textbf{I}_1^0), (\textbf{U}_2^0, \textbf{I}_2^0), \cdots ,(\textbf{U}_P^0, \textbf{I}_P^0)\}$, set of training instances $\Omega$, L2 regularization coefficient $\lambda$, number of GCN layers $L$.
\ENSURE Model parameters $\Phi$
\STATE Initialize model parameters $\Phi$ randomly
\WHILE{not converge}
\FOR{$p \in \{1, 2, \cdots, P\}$}
\FOR{$l \in \{1, 2, \dots, L - 1\}$}
\STATE Calculate $\textbf{U}_p^{l + 1}$ and $\textbf{I}_p^{l + 1}$ according to Equation (\ref{user_gcn}) and (\ref{item_gcn})
\ENDFOR
\ENDFOR
\FOR{$(u, i, j) \in \Omega$}
\STATE Obtain $\textbf{e}^u_p$, $\textbf{e}^i_p$ and $\textbf{e}^j_p$ from $\textbf{U}_p^{L}$ and $\textbf{I}_p^{L}$
\STATE Calculate $u$'s interests on $i$ and $j$ according to Equation (\ref{ewp})
\STATE Conduct interest composition according to Equation (\ref{equa_intra}), (\ref{equa_inter}) and (\ref{equation_ic})
\STATE Calculate the predicted ratings $\hat{r}_{u, i}$ and $\hat{r}_{u, j}$ according to Equation (\ref{rate})
\STATE Calculate total loss according to Equation (\ref{equa_bpr}) and (\ref{equa_loss})
\STATE Update model parameters $\Phi$ by Adam
\ENDFOR
\ENDWHILE
\RETURN Model parameters $\Phi$
\end{algorithmic}
\end{algorithm}
\subsection{Parameter Learning} \label{parameter-learning}

We construct the pairwise recommendation loss in the form of BPR according to the predicted score,
\begin{equation}
Loss_{u, i, j} = -log(sigmoid(\hat{r}_{u, i} - \hat{r}_{u, j}))
\label{equa_bpr}
\end{equation}
where $i$ denotes the positive item that user $u$ has interacted with, while $j$ is the negative item that user $u$ has not interacted with. The nonlinear function $sigmoid$ is utilized to map the value into (0, 1).

In general, the total loss is then defined as follows:
\begin{equation}
Loss_{total} = \frac{1}{\mid \Omega \mid}\sum_{(u, i, j) \in \Omega}Loss_{u, i, j} + \lambda \parallel \Phi \parallel ^ {2}
\label{equa_loss}
\end{equation}
where $\Omega$ denotes the training instances with the form of BPR in the training datasets. For each record pair $(u,i)$, we randomly sample 1 negative item $j$ to build the triple tuple in the BPR. Since the over-fitting problem may influence the training of HicRec, we adopt L2 regularization with coefficient $\lambda$, and $\Phi$ denotes all parameters in our model. Moreover, adaptive moment estimation (Adam) \cite{adam} is used to minimize the total loss and update the parameters in our model. Algorithm \ref{algorithm_hicrec} gives the overall steps of our model.

\section{Experiments} \label{experiments}

In this section, extensive experiments are conducted to evaluate the performance of our proposed model. The datasets, evaluation metrics, baselines and experimental settings are introduced first. Then, we present the experimental results and some experimental analysis.

\subsection{Datasets and Experimental Settings}
\subsubsection{Datasets}
We conducted experiments on three public datasets, Amazon, Douban movie and YELP, which all contain rich heterogeneous information. The statistics of all the datasets are summarized in Table \ref{dataset}. The Amazon dataset records users’ purchase history and the abundant auxiliary item information such as brand and category. The Douban movie dataset contains interactions between users and movies, and we include auxiliary information about movies such as actors, directors and tags. In the YELP dataset, there are users’ reviews on the business as well as the business attribute information such as the business category.
\begin{table}
\renewcommand\arraystretch{1.25}
\begin{tabular}{cccccc}
\toprule
Datasets                & Relation (A-B)    & \#Node A & \#Node B & \#A-B  & Density \\ \toprule
\multirow{3}{*}{Amazon} & User-Item         & 6170     & 2753     & 195710 & 1.152\% \\
                        & Item-Brand        & 2753     & 334      & 2753   & 0.299\% \\
                        & Item-Category     & 2753     & 22       & 5508   & 9.094\% \\ \hline
\multirow{4}{*}{Douban} & User-Movie        & 3022     & 2766     & 111471 & 1.334\% \\
                        & Movie-Actor       & 2766     & 2385     & 8418   & 0.128\% \\
                        & Movie-Director    & 2766     & 743      & 3042   & 0.148\% \\
                        & Movie-Tag         & 2766     & 31       & 6540   & 7.627\% \\ \hline
\multirow{3}{*}{YELP}   & User-Business     & 6348     & 1537     & 25169  & 0.258\% \\
                        & Business-Category & 1537     & 325      & 5084   & 1.018\% \\
                        & Business-City     & 1537     & 46       & 1537   & 2.174\% \\ \toprule
\end{tabular}
\caption{Statistics of the datasets}
\label{dataset}
\end{table}

\subsubsection{Evaluation Metrics}
Each dataset is divided into a training set and a test set. For each user, the last item of the item list the user interacted with is kept in the test set, while the others are kept in the training set. To improve the model training efficiency, we randomly sample one negative instance for each positive instance in the training set. Similar to \cite{ncf}, we also randomly select 99 items that have not been interacted with by users for testing and sorting these 100 items (1 positive and 99 negative) according to the predicted scores. Then, the {T}op-N recommendation list is obtained from the sorted list. To evaluate the performance, we adopt the Hit Ratio (HR) and the Normalized Discounted Cumulative Gain (NDCG) as the evaluation metrics,
\begin{equation}
HR = \frac{\# hits}{\# users}
\end{equation}
\begin{equation}
NDCG = \frac{1}{\# users}\sum_{i=1}^{\# users}\frac{1}{log_{2}(p_{i} + 1)}
\end{equation}
where $\#hits$ denotes the number of users who obtain correct recommendation results, and $\#users$ is the total number of users. $p_i$ is the location of the ground-truth item appears in the {T}op-N recommendation list.

\subsubsection{Baselines}
We select recent and classic RS models as baselines and their brief descriptions are listed as follows:

\noindent \textbf{MF-BPR} \cite{bpr}: Matrix Factorization is the most common collaborative filtering method. It factorizes the user-item interaction matrix into two low-rank feature matrices and makes predictions according to them. In our experiment, we construct the BPR loss function as the optimization objective of the MF.

\noindent \textbf{NeuMF} \cite{ncf}: The NeuMF is a combination of generalized matrix factorization (GMF) and MLP. Generally, among the user-item interactions, the GMF is utilized to model the linear relation, while the MLP is adopted to capture the nonlinear information.

\noindent \textbf{HNAFM} \cite{hnafm}: The HNAFM is an HIN-based RS model. It extracts information from multiple meta-paths with the help of MLP, and a two-level attention mechanism is designed to merge the information. Finally, the FM is used to predict the score given the representations of users and items.

\noindent \textbf{HueRec} \cite{huerec}: The HueRec is also an HIN-based RS model. It utilizes the GMF to predict the users’ scores for items in each meta-path and maps the user, item and meta-paths representations into a unified latent space. Then, it employs the attention mechanism to fuse the predicted scores in each meta-path into the final predicted score.

\noindent \textbf{NeuACF} \cite{neuacf}: The NeuACF splits the auxiliary information into various aspects and extracts information from them by the MLP. The user and item representations from all aspects are merged by the attention mechanism, and the predicted scores are calculated by the final merged user and item embeddings.

To illustrate the importance of the GCN and interest compositions, we also introduce two variants of our model as follows:

\noindent \textbf{HicRec$_{mlp}$}: HicRec$_{mlp}$ replaces the GCN used in the interest extraction stage with the MLP.

\noindent \textbf{HicRec$_{linear}$}: HicRec$_{linear}$ removes the interest composition in Equation \ref{rate} and only keeps the linear inner product of user/item representations.

\begin{table}[]
\renewcommand\arraystretch{1.25}
\begin{tabular}{cccc}
\toprule
Datasets                      & Aspect   & User Meta-Path {(Density)} & Item Meta-Path {(Density)} \\ \toprule
\multirow{3}{*}{Amazon}       & History  & UIU {(27.37\%)}           & IUI {(66.91\%)}           \\
                              & Brand    & UIBIU {(99.63\%)}          & IBI {(26.08\%)}           \\
                              & Category & UICIU {(100.00\%)}          & ICI {(98.41\%)}           \\ \hline
\multirow{4}{*}{Douban Movie} & History  & UIU {(62.63\%)}            & IUI {(29.33\%)}            \\
                              & Actor    & UIAIU {(97.82\%)}          & IAI {(0.77\%)}           \\
                              & Director & UIDIU {(91.08\%)}          & IDI {(0.29\%)}           \\
                              & Tag      & UITIU {(100.00\%)}          & ITI {(53.27\%)}           \\ \hline
\multirow{3}{*}{YELP}         & History  & UIU {(1.08\%)}           & IUI {(6.53\%)}           \\
                              & Category & UICaIU {(31.96\%)}        & ICaI {(15.44\%)}           \\
                              & City     & UICiIU {(80.63\%)}        & ICiI {(55.43\%)}           \\ \toprule
\end{tabular}
\caption{Aspects and corresponding meta-paths in our experiments}
\label{metapath}
\end{table}

\subsubsection{Implementation}
We initialize all the parameters of our model with an xavier initializer. The user/item embedding dimension $d$ in each meta-path is set to 32. Similarly, the latent factor dimension $K$ is also set to 32. Moreover, the coefficient of L2 regularization $\lambda$ is set to 0.0001. In the model training process, we adopt mini-batch method. The batch size and learning rate are set to 4096 and 0.001, respectively. Table \ref{metapath} shows the meta-paths and the density of corresponding graphs used in our experiments, and the same meta-paths are also used in HIN-based baselines. In the test process, the length of the recommendation list is set from 5 to 20 with step 5. For all baselines, the hyper-parameters are kept the same as the original literature except for the embedding dimension, which is set to 32. All experiments are run on a machine with a GPU (GeForce RTX2080Ti 11G).

\subsection{Experimental Results and Analysis}

\subsubsection{Overall performance}

\begin{table}
\renewcommand\arraystretch{1.25}
\renewcommand\tabcolsep{2pt}
\begin{tabular}{cccccccccc}
\toprule
Datasets                & Metrics & MF-BPR & NeuMF  & HNAFM  & HueRec & NeuACF & HicRec$_{mlp}$ & HicRec$_{linear}$ & HicRec          \\ \toprule
\multirow{8}{*}{Amazon} & HR@5    & 0.2344 & 0.2426 & 0.1830 & 0.1697 & 0.2137 & 0.2592          & 0.2508             & \textbf{0.2797} \\
                        & NDCG@5  & 0.1520 & 0.1562 & 0.1141 & 0.1015 & 0.1371 & 0.1674          & 0.1620             & \textbf{0.1827} \\
                        & HR@10   & 0.3598 & 0.3766 & 0.3004 & 0.2934 & 0.3330 & 0.4012          & 0.3957             & \textbf{0.4274} \\
                        & NDCG@10 & 0.1922 & 0.1990 & 0.1517 & 0.1407 & 0.1754 & 0.2113          & 0.2088             & \textbf{0.2303} \\
                        & HR@15   & 0.4549 & 0.4712 & 0.3863 & 0.3915 & 0.4196 & 0.5056          & 0.4994             & \textbf{0.5340} \\
                        & NDCG@15 & 0.2164 & 0.2246 & 0.1740 & 0.1665 & 0.1985 & 0.2392          & 0.2361             & \textbf{0.2588} \\
                        & HR@20   & 0.5292 & 0.5477 & 0.4604 & 0.4729 & 0.4898 & 0.5859          & 0.5808             & \textbf{0.6120} \\
                        & NDCG@20 & 0.2345 & 0.2425 & 0.1905 & 0.1862 & 0.2147 & 0.2590          & 0.2552             & \textbf{0.2751} \\ \hline
\multirow{8}{*}{Douban Movie} & HR@5    & 0.2497 & 0.2834 & 0.2638 & 0.1546 & 0.2866 & 0.2866          & 0.2889             & \textbf{0.2923} \\
                        & NDCG@5  & 0.1557 & 0.1767 & 0.1658 & 0.0935 & 0.1783 & \textbf{0.1816} & 0.1786             & 0.1805          \\
                        & HR@10   & 0.4145 & 0.4518 & 0.4309 & 0.2848 & 0.4576 & 0.4610          & 0.4582             & \textbf{0.4702} \\
                        & NDCG@10 & 0.2084 & 0.2306 & 0.2174 & 0.1347 & 0.2333 & 0.2367          & 0.2342             & \textbf{0.2379} \\
                        & HR@15   & 0.5254 & 0.5692 & 0.5532 & 0.3922 & 0.5785 & 0.5768          & 0.5805             & \textbf{0.5901} \\
                        & NDCG@15 & 0.2382 & 0.2619 & 0.2502 & 0.1632 & 0.2648 & 0.2672          & 0.2654             & \textbf{0.2686} \\
                        & HR@20   & 0.6153 & 0.6673 & 0.6444 & 0.4882 & 0.6709 & 0.6685          & 0.6725             & \textbf{0.6808} \\
                        & NDCG@20 & 0.2591 & 0.2845 & 0.2724 & 0.1860 & 0.2868 & 0.2877          & 0.2872             & \textbf{0.2900} \\ \hline
\multirow{8}{*}{YELP}   & HR@5    & 0.3300 & 0.2281 & 0.2759 & 0.1510 & 0.3447 & 0.3476          & 0.3819             & \textbf{0.3951} \\
                        & NDCG@5  & 0.2292 & 0.1506 & 0.1756 & 0.0946 & 0.2354 & 0.2336          & 0.2568             & \textbf{0.2651} \\
                        & HR@10   & 0.4488 & 0.3339 & 0.4213 & 0.2534 & 0.4985 & 0.5003          & 0.5436             & \textbf{0.5564} \\
                        & NDCG@10 & 0.2676 & 0.1851 & 0.2220 & 0.1272 & 0.2842 & 0.2830          & 0.3092             & \textbf{0.3170} \\
                        & HR@15   & 0.5279 & 0.4111 & 0.5301 & 0.3365 & 0.6138 & 0.6075          & 0.6515             & \textbf{0.6631} \\
                        & NDCG@15 & 0.2879 & 0.2047 & 0.2506 & 0.1490 & 0.3148 & 0.3113          & 0.3376             & \textbf{0.3449} \\
                        & HR@20   & 0.5904 & 0.4769 & 0.6234 & 0.4146 & 0.6958 & 0.6908          & 0.7346             & \textbf{0.7412} \\
                        & NDCG@20 & 0.3033 & 0.2210 & 0.2725 & 0.1673 & 0.3346 & 0.3308          & 0.3575             & \textbf{0.3637} \\ \toprule
\end{tabular}
\caption{Overall recommendation performance}
\label{results}
\end{table}

The overall performance of our model and the baselines is shown in Table \ref{results}, and it can be easily observed that HicRec achieves the best performance compared with all the baselines, indicating that the utilization of the GCN and interest compositions in our model is useful for recommendation. Replacing the GCN used in HicRec with MLP in HicRec$_{mlp}$ decreases the recommendation accuracy, which further proves that the GCN used in HicRec is more suitable for extracting information from meta-paths than the MLP. Specifically, the HIN-based RS model HueRec is poorer than the other baselines. In our view, the reason is probably that the GMF used in HueRec to extract information from multiple meta-paths has a weak representation ability compared with the MLP or the GCN. Moreover, HicRec$_{linear}$, which removes the interest composition, also performs worse and it confirms that the recommendation benefits from the interest composition.

However, the performance improvement on Douban dataset is limited. This
may be caused by the sparsity of the graphs extracted from the item meta-paths $IAI$ and $IDI$ on Douban dataset. As HIN-based recommendation models, including our HicRec, usually incorporating auxiliary information extracted from various meta-paths, the density of the graph extracted from a meta-path influences the recommendation performance. As shown in Table \ref{metapath}, the density of the graphs extracted from item related meta-paths IAI and IDI on Douban dataset are extremely sparse compared with the other datasets, which results in the inefficiency of auxiliary item feature extraction by GCN on the aspect of both actor ($IAI$) and director ($IDI$). While these two aspects are key to model users’ preference on movies. In this case of data sparsity, the feature extraction ability of GCN downgrades to MLP even though GCN is generally powerful than MLP in feature extraction. Thus, the performance of HicRec is sometimes worse than HicRec$_{mlp}$ on metrics such as NDCG@5.

\begin{table}[]
\renewcommand\arraystretch{1.25}
\begin{tabular}{ccccccc}
\toprule
Datasets     & BPR    & NeuMF & HNAFM & HueRec & NeuACF & HicRec \\ \toprule
Amazon       & 2.44   & 6.20  & 10.12 & 3.31   & 31.98  & 5.97   \\ \hline
Douban Movie & 0.68   & 2.31  & 4.44  & 1.29   & 15.69  & 2.75   \\ \hline
YELP         & 0.99   & 2.32  & 6.00  & 1.54   & 13.18  & 5.31   \\ \toprule
\end{tabular}
\caption{Comparison of model training time (s)}
\label{training-time-comp}
\end{table}

In addition, we also compare the model training time of our HicRec and the baseline models in the same environment with a single GPU. Table \ref{training-time-comp} shows the average training time of one epoch of each model on each dataset for the consideration of different epochs needed by different models to converge. From Table \ref{training-time-comp}, it is obvious that the training of models based on matrix factorization (BPR, HueRec and NeuMF) needs less time than those based on deep neural network (HNAFM and NeuACF). The training time of our HicRec ranges between those two types of models. The introduction of NFM in HicRec improves the recommendation performance at an acceptable training time cost. Actually, with the help of NFM in processing both the intra- and inter-meta paths, less layers are needed in GCN to extract information from meta-paths.

\begin{figure}
	\subfigure[Amazon]{
		\begin{minipage}[b]{0.95\textwidth}
		\includegraphics[scale=0.37]{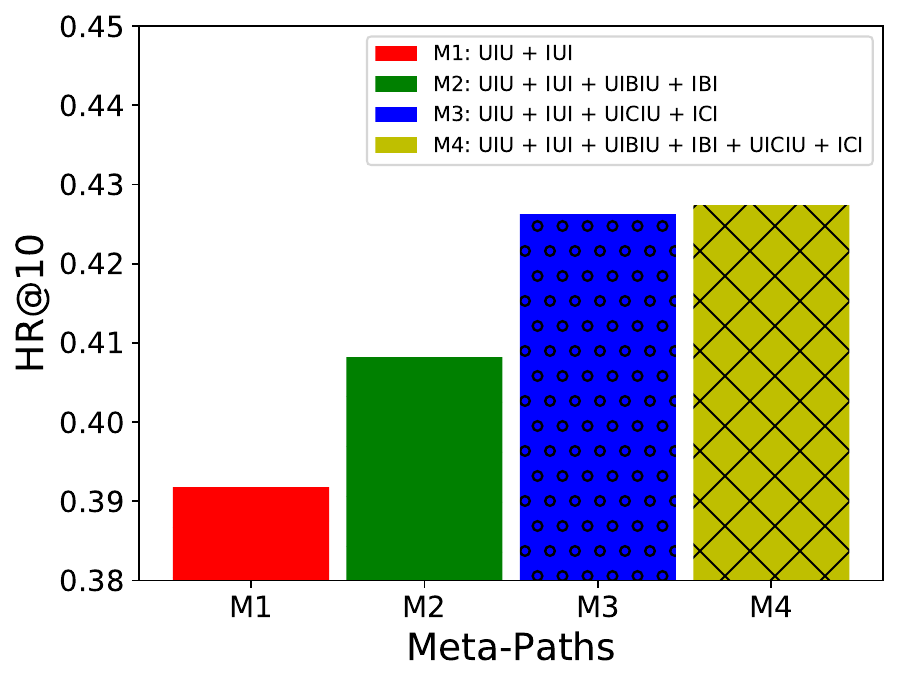}
		\includegraphics[scale=0.37]{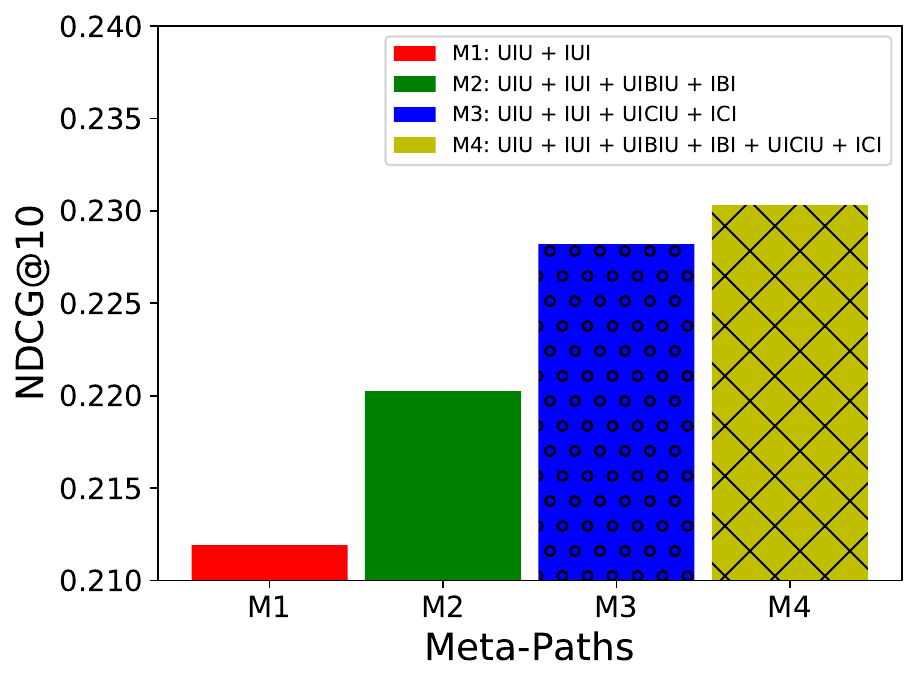}
		\end{minipage}
	}
	\subfigure[Douban Movie]{
		\begin{minipage}[b]{0.95\textwidth}
		\includegraphics[scale=0.37]{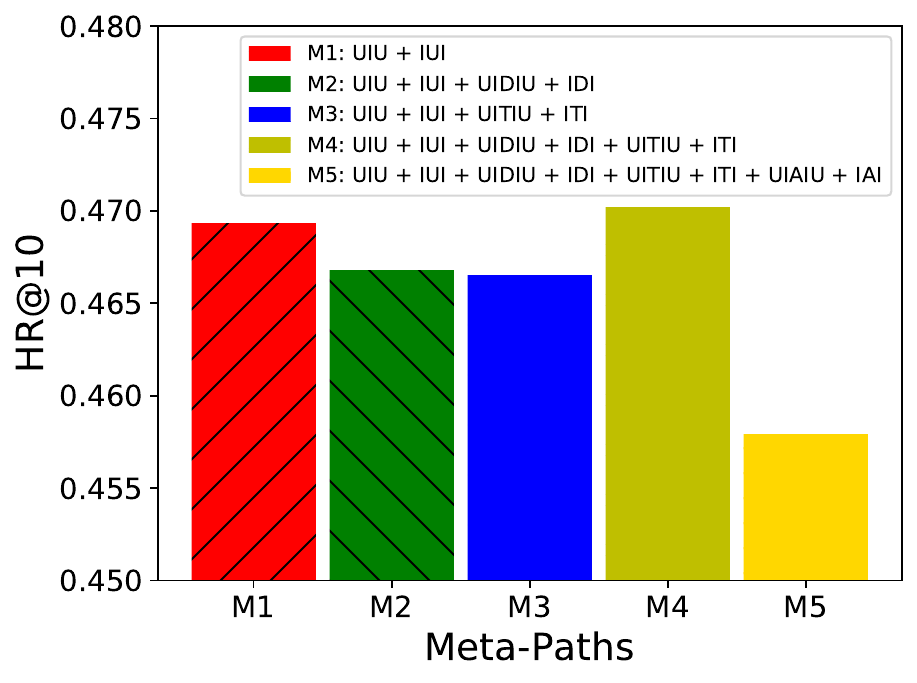}
		\includegraphics[scale=0.37]{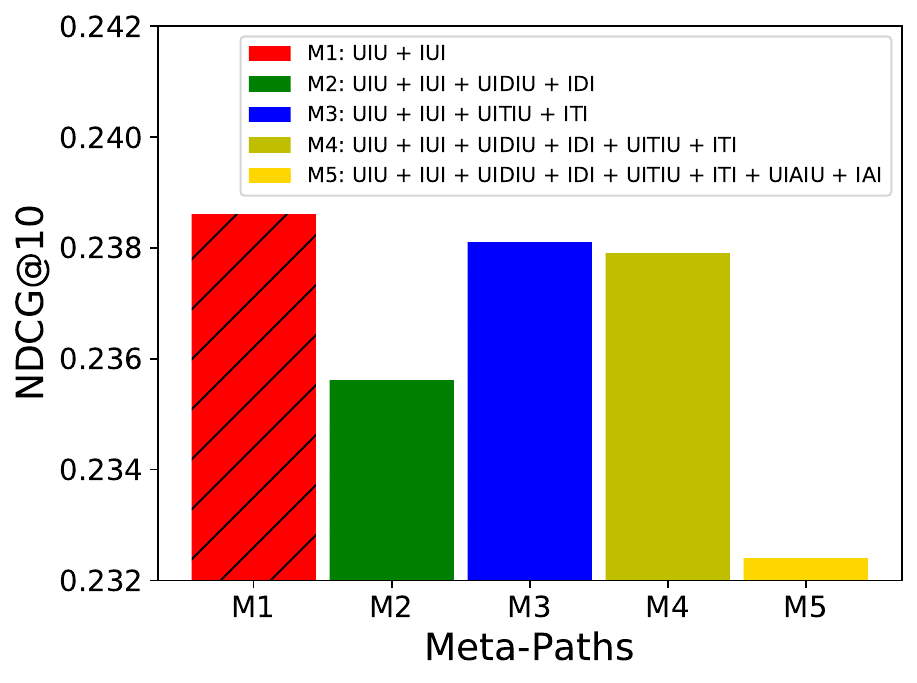}
		\end{minipage}
	}
	\subfigure[YELP]{
		\begin{minipage}[b]{0.95\textwidth}
		\includegraphics[scale=0.37]{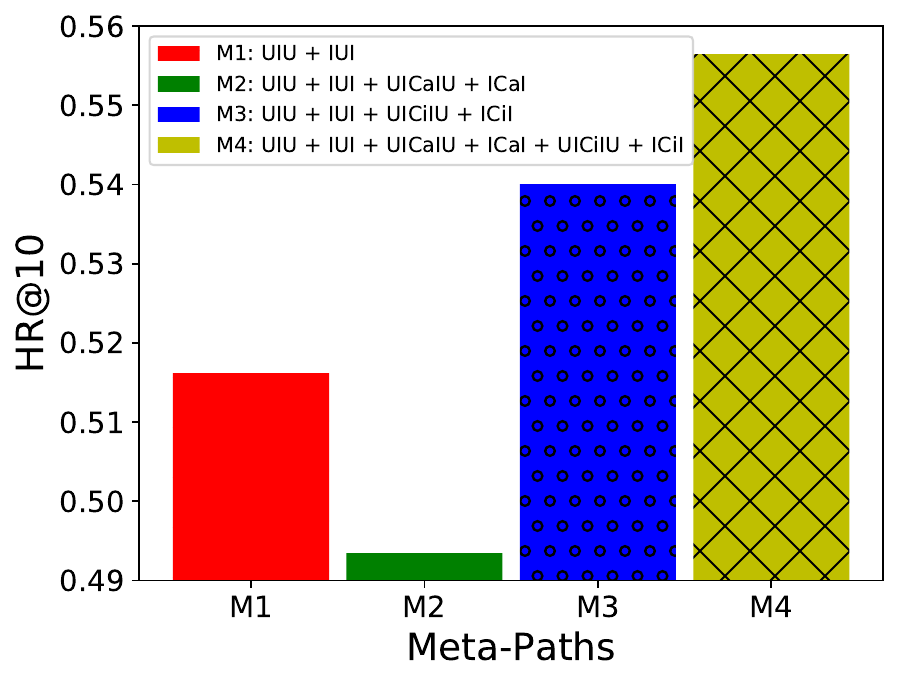}
		\includegraphics[scale=0.37]{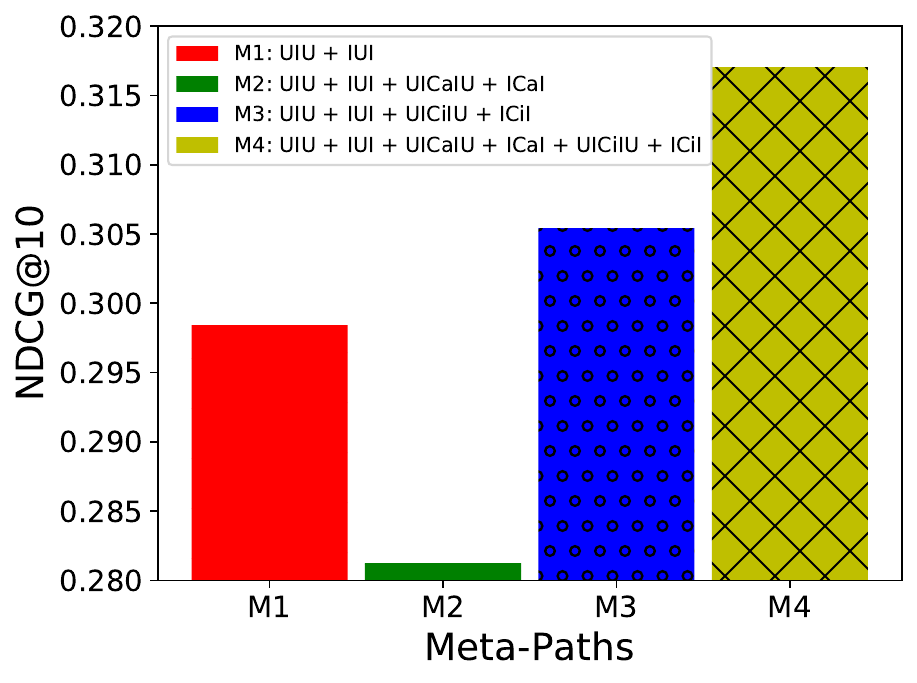}
		\end{minipage}
	}
	\caption{{Impact of different meta-path compositions for recommendation}}
	\label{result_metapath}
\end{figure}

\subsubsection{Impact of different meta-paths}
To explore the different contributions of different meta-paths to the final recommendation performance, we demonstrate the effect of various meta-paths combinations in Figure \ref{result_metapath}. Based on user-item interactions, the addition of some meta-paths can increase the recommendation accuracy (e.g., $UICIU$ + $ICI$ in the Amazon dataset, and $UICiIU$ + $ICiI$ in the YELP dataset), while the addition of some other meta-paths may decrease the recommendation performance (e.g., $UIAIU$ + $IAI$ in the Douban dataset and $UICaIU$ + $ICaI$ in the YELP dataset). Actually, the addition of meta-paths brings not only useful information but also noise, which may influence the effect of the original model. Thus, we need to search for the most suitable meta-paths in new datasets for better recommendation performance.

\begin{figure}
	\subfigure[Amazon]{
		\begin{minipage}[b]{0.95\textwidth}
		\includegraphics[scale=0.37]{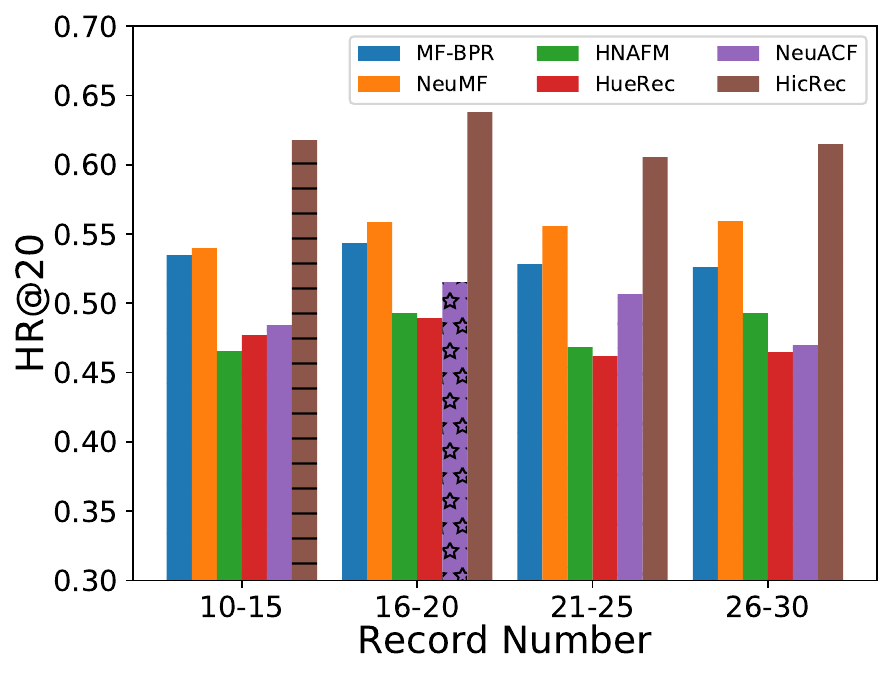}
		\includegraphics[scale=0.37]{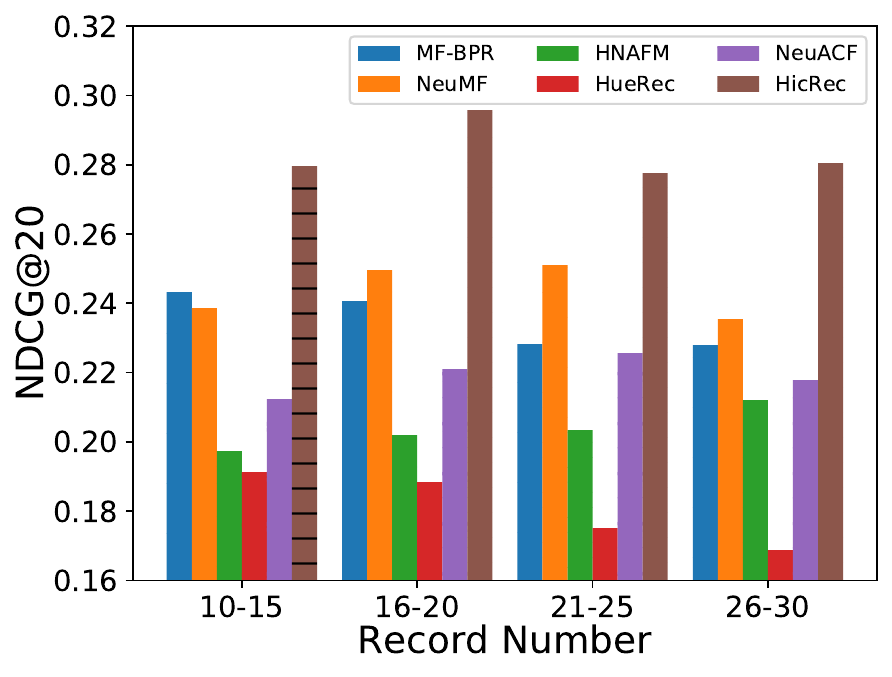}
		\end{minipage}
	}
	\subfigure[Douban Movie]{
		\begin{minipage}[b]{0.95\textwidth}
		\includegraphics[scale=0.37]{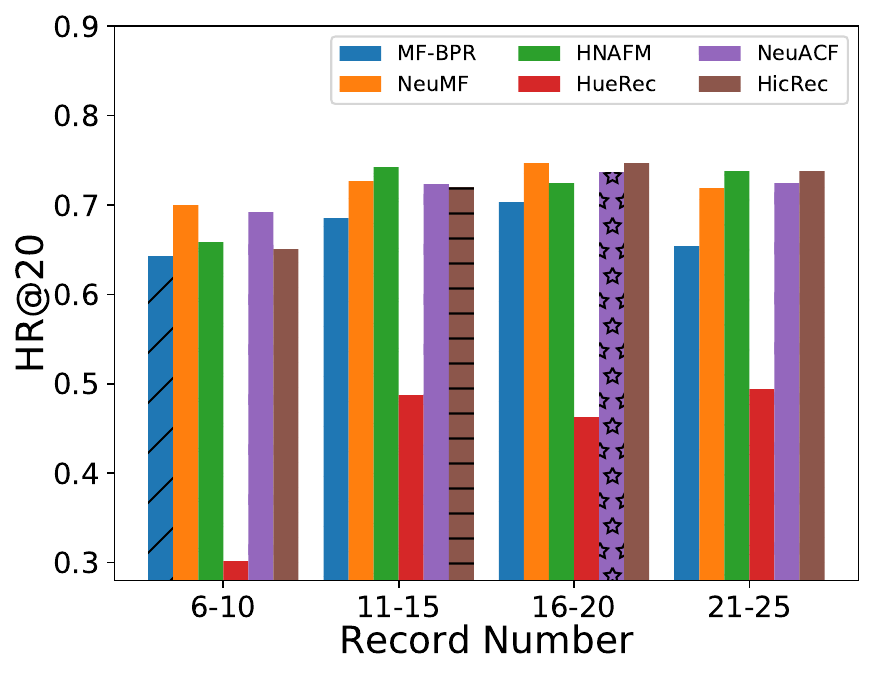}
		\includegraphics[scale=0.37]{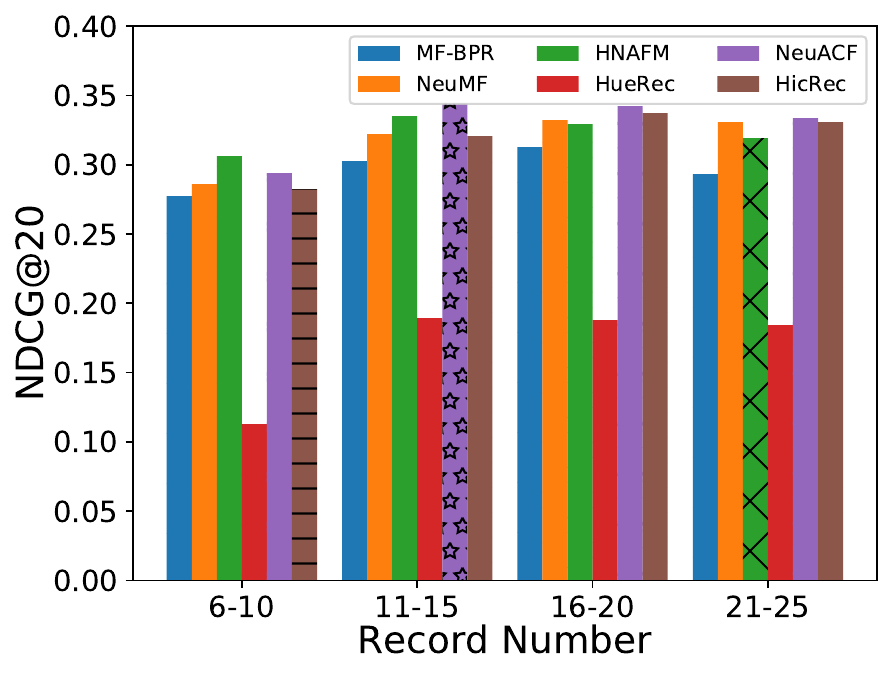}
		\end{minipage}
	}
	\subfigure[YELP]{
		\begin{minipage}[b]{0.95\textwidth}
		\includegraphics[scale=0.37]{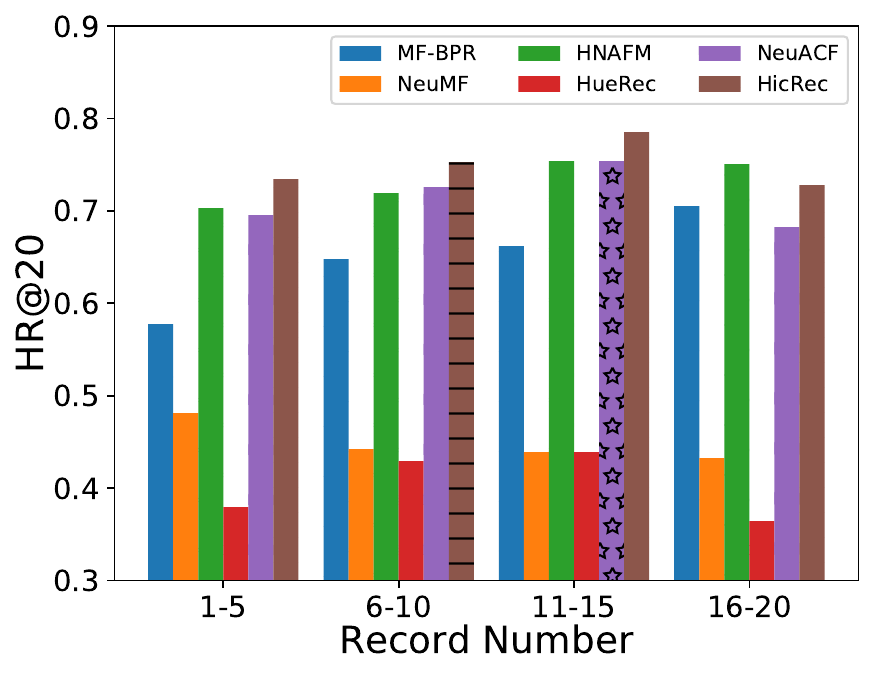}
		\includegraphics[scale=0.37]{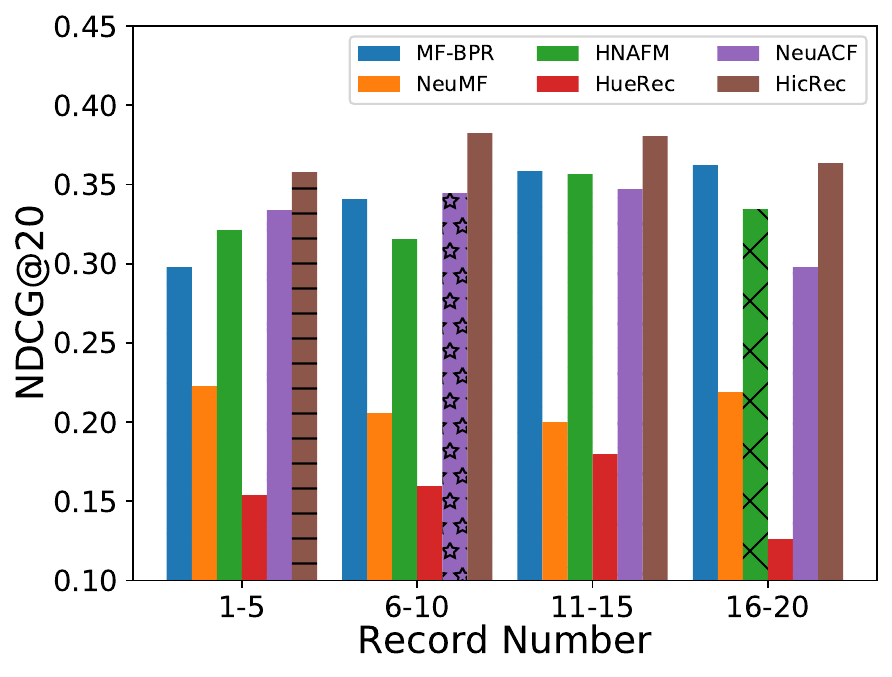}
		\end{minipage}
	}
	\caption{{Recommendation performance of different methods for cold-start users. The X-axis denotes the number of users' history records}}
	\label{cold-start}
\end{figure}
\begin{figure}
	\subfigure[Amazon]{
		\begin{minipage}[b]{0.95\textwidth}
		\includegraphics[scale=0.37]{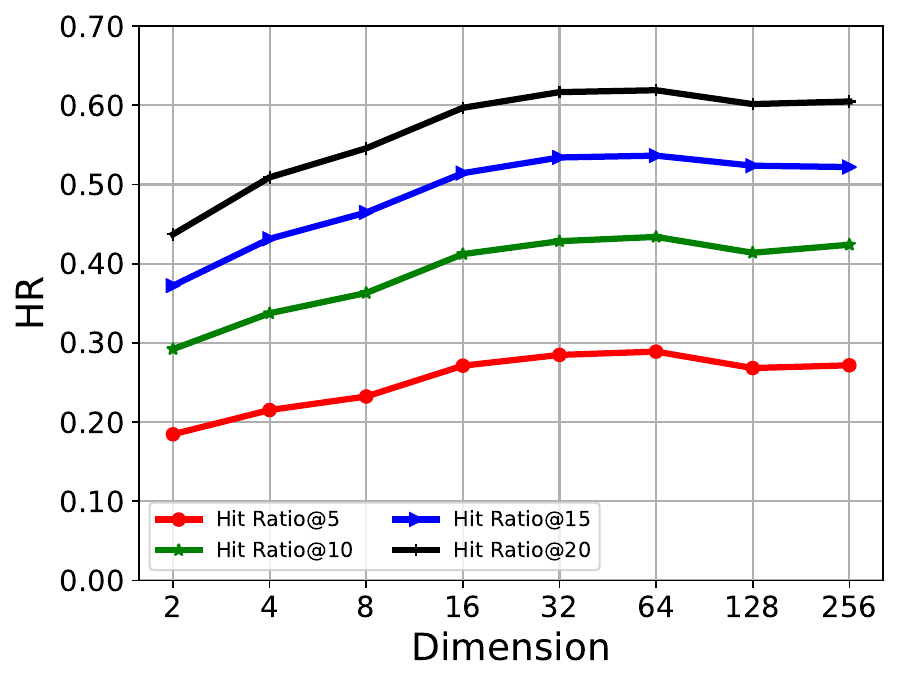}
		\includegraphics[scale=0.37]{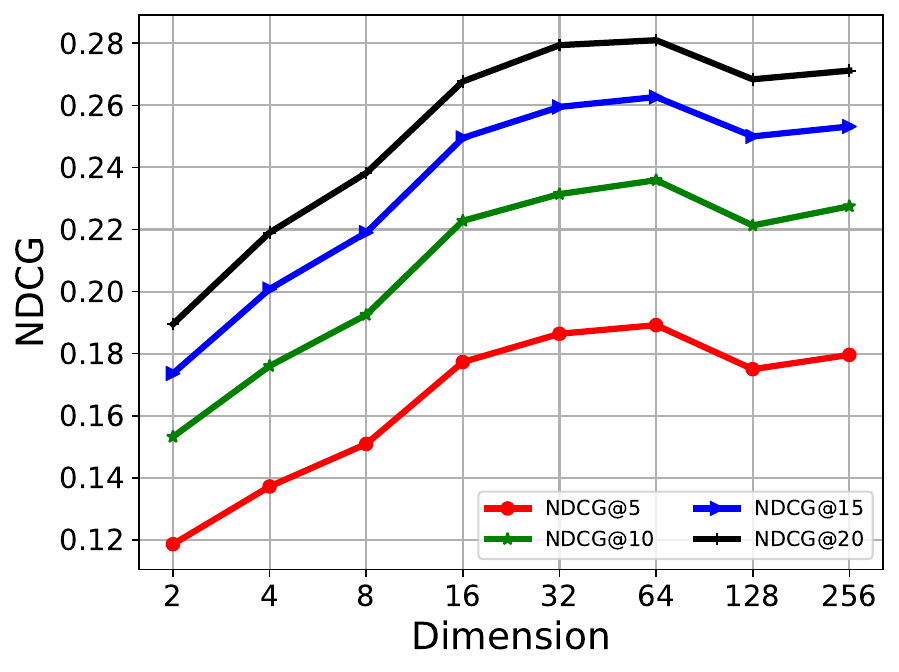}
		\end{minipage}
	}
	\subfigure[Douban Movie]{
		\begin{minipage}[b]{0.95\textwidth}
		\includegraphics[scale=0.37]{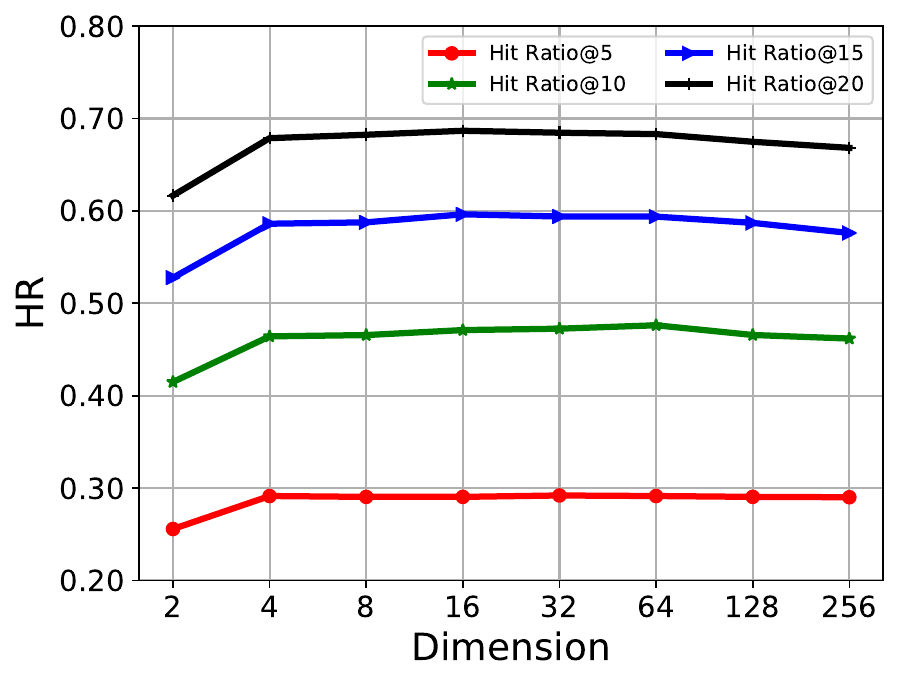}
		\includegraphics[scale=0.37]{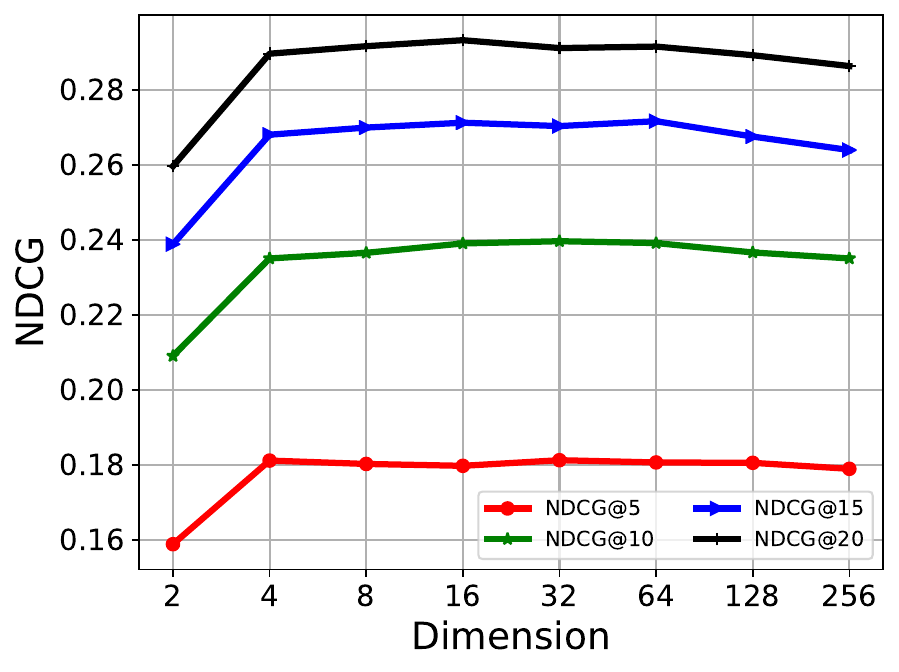}
		\end{minipage}
	}
	\subfigure[YELP]{
		\begin{minipage}[b]{0.95\textwidth}
		\includegraphics[scale=0.37]{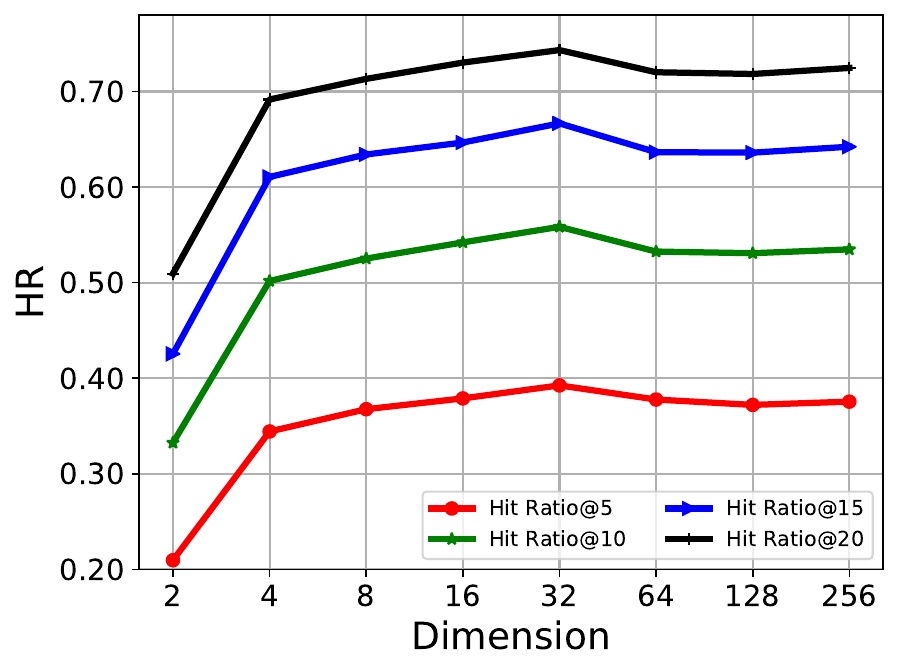}
		\includegraphics[scale=0.37]{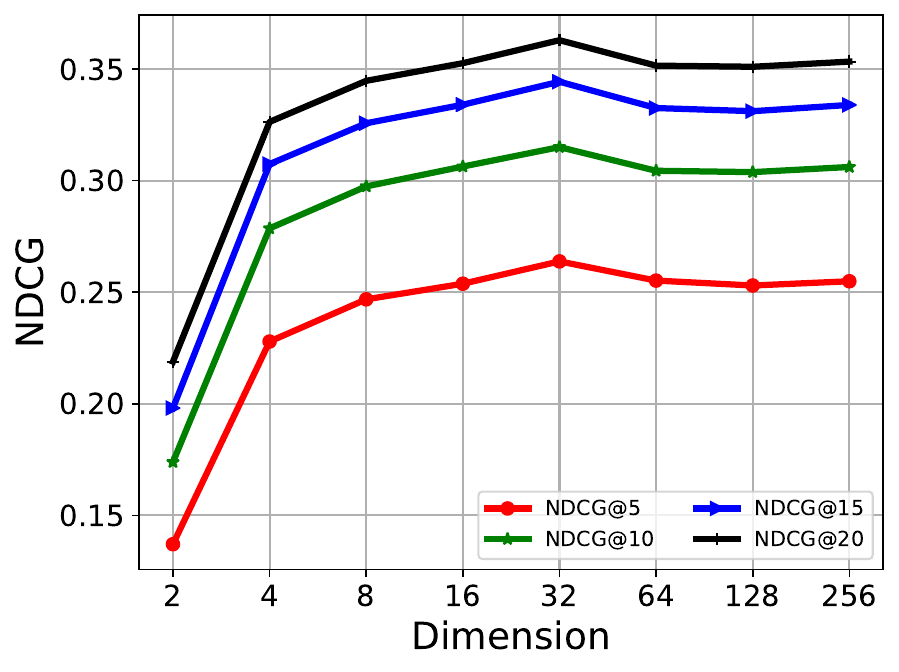}
		\end{minipage}
	}
	\caption{{Influence of embedding dimension for recommendation performance}}
	\label{dimension}
\end{figure}

\subsubsection{Recommendation for cold-start users}
{RS} usually face the challenge of cold-start and recommendation accuracy is often poor for users with fewer history records. Intuitively, the lack of records may cause {RS} to be unable to capture users’ preferences effectively. HIN-based models, including our model, employ abundant auxiliary information from the HIN to alleviate this problem. To show the effect of HicRec on this problem, we only compare the performance for cold-start users, and the results are shown in Figure \ref{cold-start}. It is obvious that HicRec can better address the cold-start problem compared with the baselines on the Amazon and YELP datasets. However, in the Douban dataset, our model cannot achieve the best performance for cold-start users. According to the dataset statistics shown in Table \ref{dataset}, the auxiliary information in the Douban dataset is sparse compared with the Amazon and YELP datasets, and we think the information extracted from these meta-paths is insufficient to capture the true preferences of these cold-start users.

\subsubsection{Impact of embedding dimension $d$}
The embedding dimension plays a vital role in its expression ability. An embedding dimension that is too low may be weak in representing complex features, while an embedding that is too high will introduce noise as well as unnecessary computational costs. Thus, the selection of the embedding dimension has a great impact on the recommendation effect. We test the impact of different embedding dimensions ranging from 2 to 256 on the recommendation performance. The results are shown in Figure \ref{dimension}. It is obvious that a dimension that is too low or too high has a negative influence on the recommendation. Moreover, the high dimensionality brings huge computation expense. The training time of our model increases almost three times when the dimension increase from 2 to 256. Therefore, a suitable embedding dimension can not only contribute well to the model effect but also save computational resources.

\section{Conclusion and Future Work} \label{conclusions}
In this paper, we propose a novel HIN-based recommendation model HicRec. With the application of the GCN and interest composition, our model significantly improves the recommendation accuracy in three public datasets. In the information extraction stage, the utilization of the GCN allows the model to learn user/item embeddings by considering not only user/item features but also the graph structure information. Compared with conventional methods that only consider either of them, the embeddings learned by our model have a more powerful representation ability. In the information fusion stage, the interest composition proposed in our model can capture users’ high-order interests intra- and inter-meta-paths. High-order interests combine users’ single interests and generate abundant derived interests. The derived interests can be used for the model to mine deeper relations among user-item interactions and improve the recommendation performance. Extensive experiments are conducted to validate the effect of our model, and the results demonstrate that HicRec outperforms baseline {RS} based on both CF and HIN. Moreover, the results of experiments for cold-start users indicate that our model is effective in alleviating the cold-start problem.

As shown in the experiments, the selection of meta-paths will significantly influence the recommendation accuracy. Indeed, the selection of meta-paths determines how much auxiliary information can be used for recommendation. Therefore, how should we seek useful meta-paths in a new dataset or scene? In future work, we will consider how to adaptively determine the meta-paths.

\section*{Ackonwledge}
This work was supported by the National Natural Science Foundation of China (No.61872002, U1936220), the University Natural Science Research Project of Anhui Province (Grant No. KJ2019A0037), and the Natural Science Foundation of Anhui Province of China (No.1808085MF197). Yiwen Zhang is the corresponding author of this paper.

%\begin{acknowledgements}
%\end{acknowledgements}


\bibliographystyle{spbasic}      % basic style, author-year citations
\bibliography{HicRec.bib}   % name your BibTeX data base

\end{document}